\documentclass[a4paper]{jpconf}
\usepackage{graphicx}
\usepackage{amsmath,amssymb}

\newcommand{\beq}{\begin{equation}}
\newcommand{\eeq}{\end{equation}}

\begin{document}

\title{Two Approaches to Dislocation Nucleation in the Supported Heteroepitaxial Equilibrium Islanding Phenomenon}

\author{J. Jalkanen$^1$, O. Trushin$^2$, K.Elder$^3,$ E. Granato$^4$, S. C. Ying$^5$ and T. Ala-Nissila$^{1,5}$}
\address{$^1$ Helsinki University of Technology, FIN-02015 TKK, Espoo, Finland}
\address{$^2$ Institute of Microelectronics and Informatics,
Academy of Sciences of Russia, Yaroslavl 150007, Russia}
\address{$^3$ Department of Physics, Oakland University, Rochester, MI, 48309-4487} 
\address{$^4$ Laborat\'orio Associado de Sensores e Materiais,
Instituto Nacional de Pesquisas Espaciais, 12245-970 S\~ao Jos\'e
dos Campos, SP Brasil}
\address{$^5$ Department of Physics, P.O. Box 1843, Brown University, Providence, RI 02912-1843}

\begin{abstract}
We study the dislocation formation in 2D nanoscopic islands with 
two methods, the Molecular Static method and the Phase Field 
Crystal method. It is found that both methods indicate the same 
qualitative stages of the nucleation process. The dislocations
nucleate at the film\--substrate contact point and the energy 
decreases monotonously when the dislocations
are farther away from the island\--wetting film contact points 
than the distance of the highest energy barrier. 
\end{abstract}

\section{Introduction}

The shape and size of islands resulting from growth processes has been a subject of numerous recent experimental and theoretical studies 
\cite{politi00,daruka99,daruka02,Moll1998,Wang2000,Wang1999,zangwill93,Johnson97,spencer01,tersoff84,Budiman,Chen1998,uemura02}. 
In particular, when the growing islands are of nanoscopic size the central issue is the possible spontaneous self-organization of islands into arrays of islands with a narrow size distribution.
Such cases offer immediate technological applications in modern nanotechnology. The details of such self-organization processes are poorly understood, however. In particular, there are still uncertainties as to whether the observed shapes and sizes of growing islands in heteroepitaxy correspond to thermodynamic equilibrium state of minimum free energy or they are limited by kinetic effects. 
In the earlier analytical and numerical studies, in addition to assuming
predefined shapes for the islands, the role of dislocation nucleation has not
been included.

In the present work we study the dislocation nucleation in two dimensional 
islands with two methods, the Molecular Static method and the Phase Field 
Crystal method (PFC) \cite{Elder04}.  The appeal of the latter method is that we do not need to 
initialize the transition path by hand. The way in which the two models reach 
the timescale of the nucleation process is quite different and our focus is on the qualitative 
stages of the process and on the agreement between the two approaches.  

In the first section on this paper we explain the methods. We use the first, 
atomistic, model to study the strain relief caused by dislocation formation 
by minimizing a transition path between the coherent and dislocated states 
with Nudged Elastic Band (NEB) Method \cite{Jonsson}. Then we explore the same 
process with the PFC method in two dimensions. The nucleation is seen 
to be qualitatively similar with both approaches. 

\section{Models and Methods}
\subsection{Molecular Static Method}
We adopt the 2D model used previously in Ref. \cite{jalkanen05}, 
Interactions between all atoms in the system are described by a modified 
Lennard-Jones (LJ) pair potential \cite{zhen83} $V(r)$ with two parameters, 
namely the dissociation energy $\varepsilon_{\rm ab}$ and the atomic equilibrium distance $r_{\rm ab},$ 
\begin{equation}
\label{LJpot}
V_{ab}(r) = \varepsilon_{\rm ab}\left[\frac{5}{3}\left(\frac{r_{\rm ab}}{r}\right)^8 -  
\frac{8}{3}\left(\frac{r_{\rm ab}}{r}\right)^5\right]
\end{equation}
with a smooth cut\--off as in \cite{zhen83}.
The indices $\rm ab$ of the equilibrium distance are $\rm ss$, $\rm ff$ and $\rm fs$ for the substrate-substrate, adsorbate-adsorbate and adsorbate-substrate interactions, respectively. 
We model the attraction and lattice mismatch between the two materials by changing the potential depth $\varepsilon_{\rm ab}$ and the equilibrium interatomic distance $r_{\rm ab}$ between the substrate and the film. 
The substrate-substrate interaction parameters 
were set to $\varepsilon_{\rm ss} = 3410.1$ K and $r_{ab}=2.5478$ {\AA}
corresponding to a Cu substrate \cite{zhen83}. 
In terms of the lattice mismatch $r_{\rm ff} = (1+f) r_{\rm ss}$ and 
$r_{\rm fs}=(r_{\rm ff}+r_{\rm ss})/2$. 
A positive mismatch $f>0$ corresponds to compressive strain and
negative $f<0$ to tensile strain when the adsorbate island is
coherent with the substrate. 
Analogously we define an interaction parameter 
$ \kappa = (\varepsilon_{ss} - \varepsilon_{sf})/\varepsilon_{ss}$
which describes the relative difference between the potential minimum 
depths in the substrate-adsorbate and absorbate-adsorbate interactions.
A negative value of $\kappa<0$ corresponds
to an effectively attractive and positive $\kappa>0$ to a repulsive substrate.
The potential depths 
can be written as $\varepsilon_{\rm ff} = \varepsilon_{\rm ss}$,  
$\varepsilon_{\rm fs} = (1+\kappa)\varepsilon_{\rm ss}$
for the adsorbate-adsorbate and substrate-adsorbate interactions, respectively.
We used periodic boundary conditions in the horizontal direction. 
Two bottom layers of the substrate were held fixed to 
simulate a se\-mi-in\-fi\-ni\-te substrate while all other layers were free to relax.  
\subsection{Phase Field Crystal Method}
\begin{figure}[t]
\includegraphics[width=19pc]{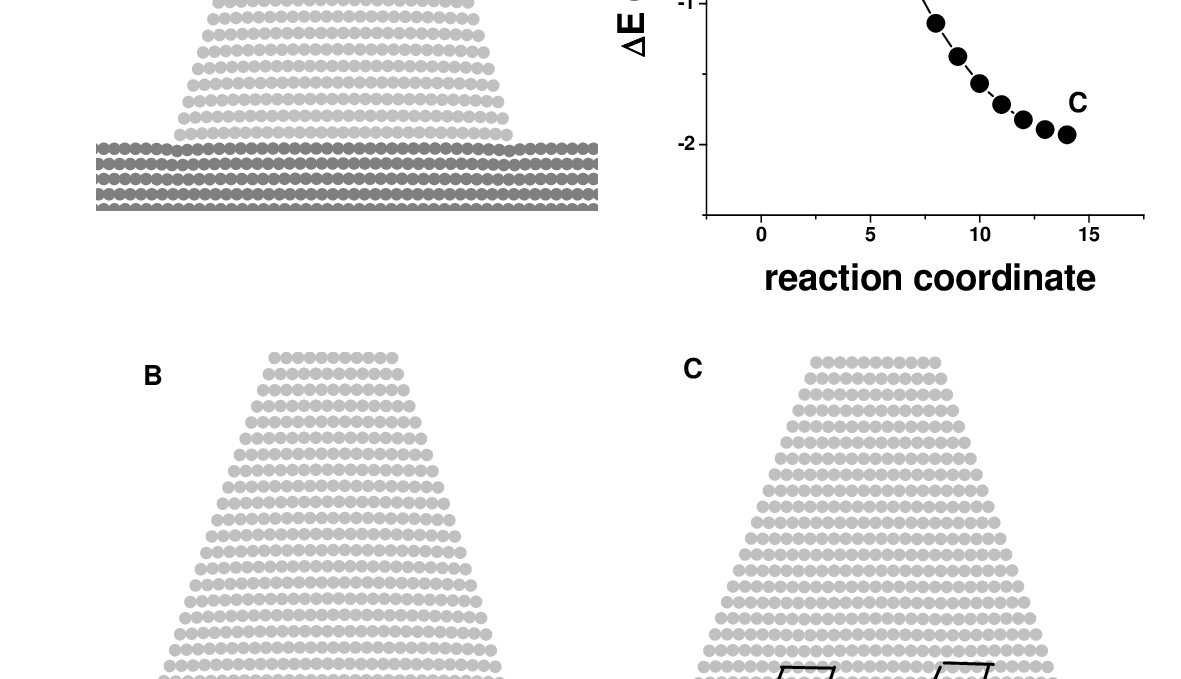}\hspace{1pc}
\vspace*{2mm}
\begin{minipage}[b]{17pc}\caption{\label{ljnebfig}
The minimum energy path between coherent and dislocated configurations
is obtained with Nudged Elastic Band method in system with a 411\-- atom island and 5 layers thick wetting film \cite{jalkanen05}. The dislocations are nucleated at the island\--wetting film contact points. At a certain distance from the 
contact point the energy barrier is highest. After this point the 
energy is lowered monotonously as the dislocations move to the island centre.}
\end{minipage}
\end{figure}

\begin{figure}[t]
\begin{minipage}[b]{38pc}
\includegraphics[width=12.3pc]{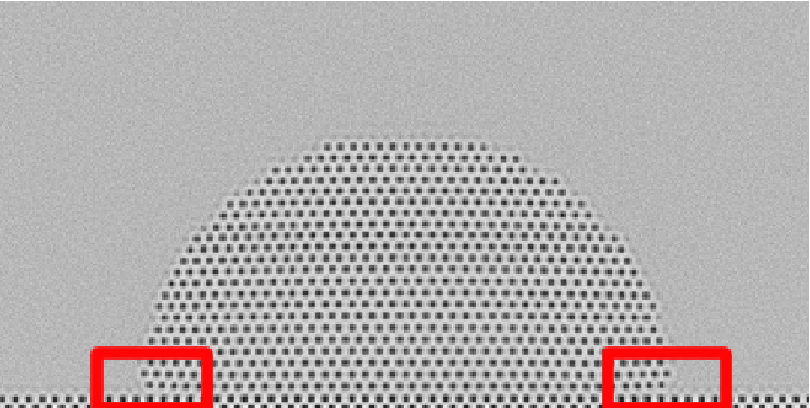}\hspace{0.6pc}%
\includegraphics[width=12.3pc]{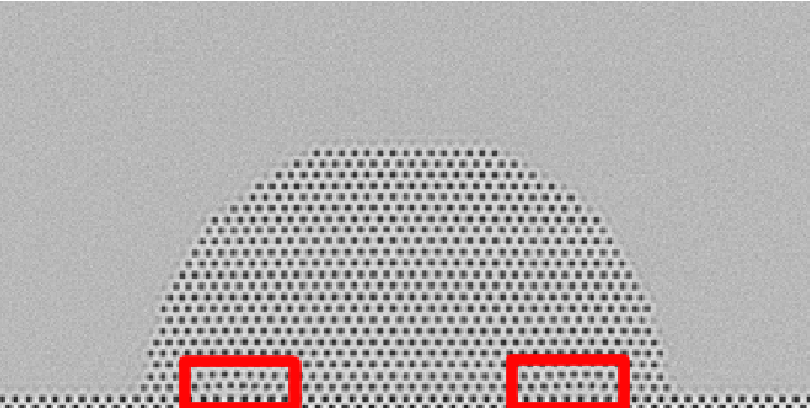} \hspace{0.3pc}%
\includegraphics[width=12.3pc]{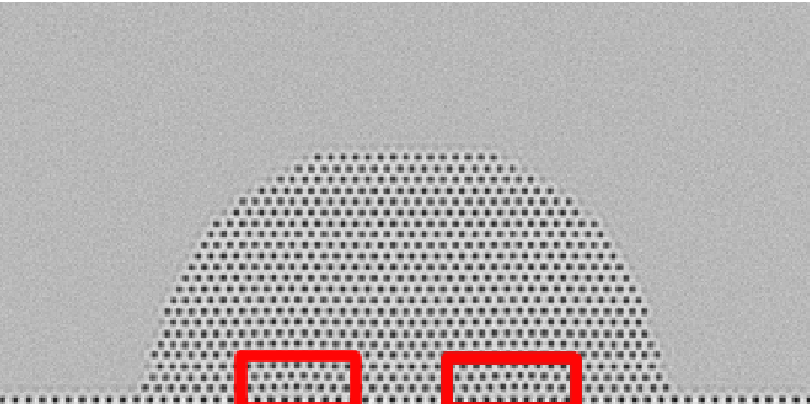}
\end{minipage}\\
\begin{minipage}[b]{17pc}
\includegraphics[width=17pc]{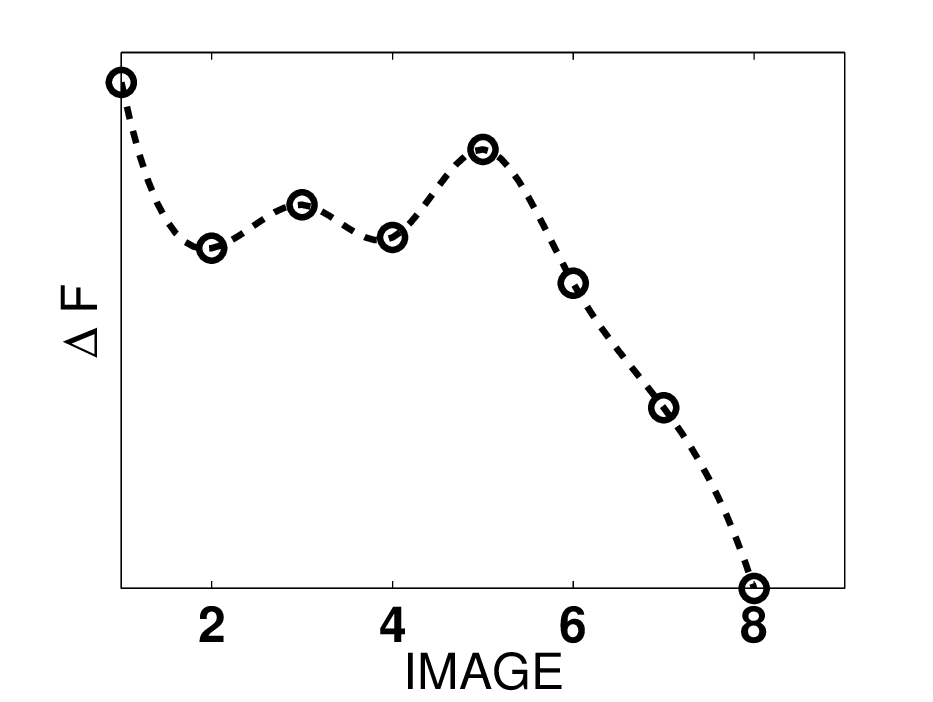}
\end{minipage}
\begin{minipage}[b]{19pc}
\caption{\label{PFCfig} The figures in the upper row
are PFC configurations at different stages of dislocation formation.
The figure in the left shows the free energy landscape as a 
function of the image number in arbitrary units. The figures in the 
upper row correspond to images 1, 5 and 8. The misfit $f = -7.8 \%$ and 
the pinning potential depth is half of the amplitude in the 
periodic phase. The line in the energy landscape is 
drawn to guide the eye. The dislocations positions are indicated by the 
square loops.}
\end{minipage}
\end{figure}

The phase field crystal method (PFC) has been applied in a variety of problems related to non\--ideal crystal structures \cite{Elder04}. The method finds
the ground state of a given truncated free energy functional
without any reference to the instantaneous position of individual atoms.
This enables us to find the dislocated ground states even if they were
behind kinetic barriers, which would be insurmountable for 
atomistic simulations.
The free energy is 
\beq
\label{thefreeenergy}
F = \int d\vec{r}\left\{\frac{1}{2}\varrho
\left[\vartheta + \left(q^2 + \nabla^2\right)^2\right]
\varrho + \frac{1}{4}\varrho^4 + h\varrho\right\}
\eeq
and the parameters $\vartheta = 1/2$ and $q = k(1 + f/100),$ while $k$ is the 
period of the pinning potential to be defined later. The components of the vector $\vec{r} = (x,z).$ We use periodic boundary conditions in 
the horizontal and mirror boundary conditions in the vertical direction. 
The function $h$ is a pinning potential, which zero, if $2 z > 3 \sqrt{3} \pi/ k$ 
and otherwise 
\beq
h(\vec{r}) = \bar{h}\left[\cos(k x) \cos(k z/\sqrt{3}) - \cos(2 k z/\sqrt{3})/2\right],
\eeq
where $k = b\sqrt{3}/2.$ The factor $b \sim 1$ ensures that the potential is 
continuous in the horizontal direction. The amplitude $\bar{h}$ is the pinning potential 
depth and it plays the same role as the interaction parameter in the 
Molecular Static method. The difference of the pinning potential and 
triangular phase periods $k$ and $q$ is adjusted to give the wanted 
lattice mismatch.
The time step $\Delta t = 0.005$ and the grid size $\Delta x = 0.7854.$
The simulation was done on $1024 \Delta x \times 512 \Delta x$ grid.
The average density $\bar{\varrho} = 0.37.$ 
This corresponds to a point in the middle of the constant phase\--triangular phase coexistence
in the one mode approximation of the phase diagram of Eq. \ref{thefreeenergy}.
The initial state has an island of half\--disc shaped region of 
triangular phase on the bottom edge of the simulation cell, see Fig.\ref{PFCfig}. This island is surrounded by constant phase. 
The initial state is evolved according to
\beq
\frac{\partial \varrho}{\partial t} = \nabla^2\frac{\delta F}{\delta \varrho} + \zeta,
\eeq
where $\zeta$ is Gaussian distributed conserved noise \cite{Elder04}.

\section{Results and Discussion}

Fig. \ref{ljnebfig} shows an energy path between coherent and dislocated  configurations \cite{jalkanen05}. 
The misfit $f = 8\%$ and $\kappa = 0.$ 
The initial chain of states is obtained by simple linear interpolation. 
This chain is minimized with NEB method 
\cite{Jonsson}. The results show an energy barrier for the initial detachment of the dislocation from the island\--wetting film contact point. 
When the distance to the contact is larger than the distance of the 
highest energy peak, the energy decreases monotonously as the dislocations approach the island center. Finally the defects stop at a finite distance from each other. 
 
We used the PFC model to make a similar calculation. We used
mismatches of $\pm 7.8 \%$ and pinning potential depth $0.5$ 
times the amplitude of the triangular phase.
The initial state is the upper half of a disk standing on the 
simulation cell edge where the pinning potential is effective.
For compressive mismatch, the island does not form dislocations 
but adopts a concave shape with reduced size of the island\--wetting film
contact area. 
 
In the tensile case dislocations are nucleated symmetrically
from both sides of the island base. This is shown in Fig. \ref{PFCfig}. 
The defects move towards the island centre. The free energy landscape 
is given in arbitrary units as a function of the image number.
It shows the combined effect of dislocation nucleation and 
facet formation. These effects could be studied separately by 
starting from dislocated or faceted initial configuration. 
The general shape of the curve in Fig. \ref{PFCfig} 
is similar to that of Fig. \ref{ljnebfig}.
 
The island shape is almost rounded in the compressive and strongly 
faceted in the tensile case. The asymmetry of tensile and 
compressive has been studied also in \cite{tru02a,tru02b}.
This asymmetry is clearly visible in the PFC model, because 
it does not nucleate dislocations in the latter case. 

\section{Conclusions}
We have studied dislocation nucleation in 2D strained heteroepitaxial nanoislands. We have employed both an atomistic model and phase field crystal method.
The results are qualitatively consistent with each other. The dislocations nucleate near the island\--wetting film contact under tensile strain. After an initial energy barrier the dislocation detaches from the contact point and moves to the middle of the island. During the move the energy is lowered gradually. The island shapes in the tensile case were observed to be more faceted than in the compressive case.

\section{Acknowledgements} 
J. Jalkanen has also been supported by the Foundation Vilho, Yrj\"{o} ja Kalle V\"{a}is\"{a}l\"{a}n rahasto of Finnish Academy of Science and Letters, a joint fund under EU STREP 016447 MagDot and NSF (DMR-0502737) and the Academy of Finland through its Center of Excellence Grant (COMP).

\end{document}